\documentstyle[prd,aps,amsmath,epsf,graphicx]{revtex}

\DeclareFontFamily{OT1}{rsfs10}{}
\DeclareFontShape{OT1}{rsfs10}{m}{n}{ <-> rsfs10 }{}
\DeclareMathAlphabet{\mathscript}{OT1}{rsfs10}{m}{n}

\newcommand{\be}{\begin{equation}}
\newcommand{\ee}{\end{equation}}

\newcommand{\bea}{\begin{eqnarray}}
\newcommand{\eea}{\end{eqnarray}}
\newcommand{\ba}{\begin{array}}
\newcommand{\ea}{\end{array}}

\newcommand{\ns}{\normalsize}

\def\a{\alpha}
\def\b{\beta}

\def\e{\epsilon}

\def\k{\kappa}

\def\cV{{\cal V}}

\begin{document}

\begin{titlepage}

\title{
\hfill{\ns DCPT-04/23\\}
\hfill{\ns hep-th/0406241\\[2cm]}
{\Large An explicit example of a moduli driven phase transition in heterotic models.}\\[1cm]}
\setcounter{footnote}{0}
\author{{\ns\large
  James Gray$^1$\footnote{email: J.A.Gray2@durham.ac.uk}
 \\[1cm]}
      {\ns $^1$ Department of Mathematical Sciences - CPT , University of Durham,}\\
      {\ns Science Laboratories, South Road, Durham, DH1 3LE, UK. }\\[0.2cm]
     }


\maketitle

\vspace{2cm}

\begin{abstract}
We present an explicit example of a gauge symmetry breaking phase transition in heterotic 
models, the dynamics of which are not 
thermal and can be described in a well controlled manner throughout. The phase transition is 
driven by the evolution of bundle moduli - moduli associated with gauge field vacuum expectation 
values in the hidden dimensions. We present the necessary parts of the four dimensional effective 
theory including moduli which describe the embedding of the gauge bundle within 
the gauge group. We then present exact cosmological solutions to the system before going on to use them to
describe the phase transition. The explicit nature of our description enables us to plot how the gauge 
bosons associated with the symmetries which are broken in the transition gain masses with time. 
This is in contrast to the use, for example, of a brane collision as a modulus driven phase transition.
In the course of this work we find a number of other interesting results. We observe that the K\"ahler potential 
of the system is given by the logarithm of the volume of the compact space even when bundle moduli are included. 
We also note that the dynamics of the gauge bundle mean that small instanton transitions are classically 
forbidden for all but a set of measure zero of the initial conditions of the system. The paper is written in 
such a manner that the cosmological description of the phase transition can be read independently of the 
derivation of the four dimensional theory.
\end{abstract}

\thispagestyle{empty}

\end{titlepage}

\renewcommand{\thefootnote}{\arabic{footnote}}


\section{Introduction}
\label{intro}

The study of four dimensional $N=1$ supersymmetric 
vacua of M-theory has two main goals from a phenomenological point of view.
The first is to find vacua which 
reproduce those features of the four dimensional effective theory describing our 
universe which have already been observed. For example such models should aim to reproduce the 
low energy particle physics described by the standard model. The second is that studying 
phenomenology in this wider context may allow us to investigate qualitatively new phenomena which 
we have not thought to look at before.

An example of the latter is given by
 what we will call 'moduli driven phase transitions'. In a standard thermal 
phase transition the mass scale associated with the phase transition, which determines the mass of
any particles which become heavy during the transition, is associated with the temperature at 
which the change of phase occurs. A moduli driven phase transition by contrast occurs, as the 
name suggests, when a modulus field evolves (before moduli stabilisation)
to achieve a certain 
value. The time at which this happens, and in particular the temperature of the universe at this 
point, is a function of the initial conditions of the system. As such the link seen in thermal 
phase transitions between temperature and the mass scale of the transition is lost.

One example of such a phase transition is gaugino condensation \cite{Dine:rz,Horava:1996vs,Lukas:1997rb} 
 in heterotic models 
\cite{Green:mn,Horava:1995qa,Horava:1996ma,Witten:1996mz,Lukas:1997fg}.
 Here a supersymmetry breaking phase 
transition can be initiated when the dilaton evolves so as to achieve a value where a gauge interaction 
becomes strongly coupled. Another 
example is a small instanton transition \cite{Witten:1995gx,Ganor:1996mu}.
In a heterotic model it is thought that 
such a transition can be initiated when an M5 brane moving in the orbifold 
direction of heterotic M-theory (in the strongly coupled language) collides with one of the 
orbifold fixed planes. Upon collision the M5 brane can disappear to be replaced with a solitonic 
gauge field configuration, dressed up with various other fields, living on the fixed plane. This is
a phase transition in the sense that
the gauge five brane (the solitonic object) which appears will in general break some of the gauge 
symmetry previously observed in the four dimensional theory. The time at which this phase transition, or 
small instanton transition, occurs is not determined by the temperature observed in the four 
dimensional cosmology associated with the system but rather by the choice of 
initial conditions. In fact we 
can choose this phase transition to occur at any time by a suitable choice of integration constants
in the solutions describing the moving M5 brane \cite{Copeland:2001zp}.

This example of a modulus driven phase transition was used in \cite{Bastero-Gil:2002hs} 
as the basis for a scenario 
of baryogenesis. The idea of that paper was to use a late time phase transition to trigger 
baryogenesis. In this way the authors could choose the temperature of the universe to be very small
as compared to the mass of the particles which were decaying to give the relevant asymmetry. As such 
'wash out' scattering processes, which can 
destroy baryon asymmetry created in such scenarios, could be completely
turned off. This allowed the scenario to obtain a reasonable value for the baryon asymmetry for a much 
wider range of parameters in the low energy particle physics model than is normally 
possible.

The problem with the examples given above as
 moduli driven phase transitions is that the 
transitions themselves are poorly understood. One can not describe 
what is actually happening during the phase transition and must resort to merely parameterising it.
The purpose of this paper is to provide an explicit example of a moduli driven phase transition in 
a heterotic compactification of M-theory which can be described in a well controlled way 
throughout. For example we would like to be able to plot how the mass of gauge bosons which are 
becoming heavy during the phase transition change with time.

We shall achieve this by considering a situation where the phase transition is driven by the 
evolution of bundle moduli, moduli which describe the evolution of the vacuum expectation value 
of the gauge fields 
\cite{Buchbinder:wz,Buchbinder:2002pr,Buchbinder:2002ic,Buchbinder:2002ji,He:2003tj,Gray:2003vw}.

In particular we shall present exact solutions to the four dimensional effective theories associated 
with certain compactifications of heterotic models which describe the gauge bundle evolving in the 
following manner. The various moduli of the heterotic model will start off evolving in the 
usual, 'freely rolling radius', manner 
with the moduli describing the gauge bundle being constant. Then at some time, which 
shall be determined by a choice of integration constant, the bundle moduli will quickly change 
from their initial to some final constant values. In doing so the little group of the bundle within 
$E8 \times E8$ (or $SO(32)$) will change and thus so will 
the visible gauge symmetry. This then will be our gauge symmetry changing 
phase transition. 

We shall be able to describe this process in a well controlled way throughout, in stark contrast 
to the case of brane collision mentioned above. In particular we shall be able to present plots 
showing how the mass of the gauge bosons that are associated with parts of the gauge symmetry 
broken in the transition change with time.

To achieve this we shall need some method for describing the gauge bundle associated 
with the vacuum solution of our compactification and in particular for deriving the four 
dimensional effective theory which includes the relevant bundle moduli. We shall use the formalism 
developed in \cite{Gray:2003vw} to explicitly describe the gauge field vacuum expectation values. For 
the application at hand we shall require a more complete description of the four dimensional 
effective theory than was presented in \cite{Gray:2003vw} 
and so before we can go on to study the phase 
transition, which is the focus of this paper, we shall first have to derive the appropriate four 
dimensional effective theory.

The specific situation we shall describe is an $SU(2)$ valued 
instanton living on a Calabi-Yau threefold internal space as part 
of a larger gauge bundle. We shall include in our analysis 
moduli which describe the embedding of this gauge field configuration within the overall gauge group.
Before the phase transition the instanton will be embedded within the gauge group so as to not 
break any gauge symmetry in the low energy theory that is not already broken by the rest of the bundle.
Then during the phase transition one of the instanton's embedding moduli will evolve to a new value 
before becoming constant once more. This change in the instanton's embedding will cause it 
to break a different part of the gauge symmetry in the low energy theory - part that was unbroken 
before the phase transition.

\vspace{0.2cm}

In the course of this analysis we shall obtain some interesting results in slightly different directions. 
We shall show that, even when the bundle moduli are included, the K\"ahler potential is still 
given by the logarithm of the volume of the compact manifold in these Calabi-Yau reductions.
We shall also see that in generic situations, where several of our bundle moduli are evolving, small 
instanton transitions are classically forbidden for all but a set of measure zero of the initial conditions.
By classically forbidden we mean that the system can not classically evolve to the singular point in the 
moduli space associated with the small instanton transition.

\vspace{0.2cm}

The outline of this paper is as follows. In the next section we will derive the four dimensional 
effective theory describing a class of heterotic models 
that include the embedding moduli described above. Section 
III will contain our description of the cosmological evolution that follows from this theory and, 
in particular, an explicit description of the moduli driven phase transition which is the goal 
of this paper. We will conclude 
in section IV. The paper is written so that the reader who is interested in the 
description of the phase transition, but not in the derivation of the four dimensional theory we 
shall use, can skip straight to section III. An understanding of section II is not required for 
reading the rest of the paper.


\section{Theory of embedding moduli.}
\label{theory}

\subsection{Overview: Explicit constructions of gauge bundles on orbifold compactifications 
of heterotic theories.}
\label{method}

We are going to be working with an orbifold reduction to four 
dimensions of the 
ten dimensional effective description of a heterotic model \cite{Green:mn,Lukas:1998ew}.
The fixed points of the orbifold will 
be blown up to give a smooth internal space. Due to its central role in our 
discussion we require an explicit description of the gauge field configuration in this 
hidden space. We shall obtain this description using the method introduced in 
\cite{Gray:2003vw}.
If the four dimensional theory associated with our dimensional reduction is to have an $N=1$ 
supersymmetric Minkowski space vacuum then the orbifold must have $SU(3)$ holonomy (at zeroth order in $\a'$) 
and the gauge field must be 
a solution to the Hermitian Yang-Mills equations.
Self-dual gauge field configurations are one type of solution to the Hermitian Yang-Mills equations 
\cite{Strominger:et} and so we shall
look for such a configuration living on our orbifold. 

 The basic idea is very simple. The portions of the orbifold compactification which 
are away from the (resolved) fixed points simply look like six dimensional euclidean flat space. 
The ADHM construction provides us with 
 a powerful formalism for constructing self-dual gauge field configurations in four 
dimensional flat space. To include the gauge field configuration on our orbifold then we just 
take a localised self-dual gauge field configuration in four dimensions - as provided by the 
ADHM construction, and take this to be sitting on our orbifold away from the fixed points. 
Four of the orbifold directions can be taken to be the spatial directions which appear in the 
ADHM solution and the configuration is taken to be independent of the remaining torus - i.e it 
is extended in these directions. All 
that remains then is to include the back reaction of the gauge field on the ten dimensional 
metric and dilaton as described in \cite{Strominger:et}. Once the self-dual gauge field is 
known this backreaction can be calculated uniquely up to discrete choices. Thus we end up 
with the picture of a localised gauge field configuration, supplied to us by the ADHM construction, 
dressed up with various other fields and located on one of the flat portions of the orbifold. 
This construction provides a good approximation to an exact solution of the appropriate killing
spinor equations near to the core of our localised gauge field configuration. 
The details of this well controlled approximation are described at 
some length in \cite{Gray:2003vw}. 

We can make our vacuum richer in structure by adding other components. For example we can 
clearly add Wilson lines to our picture without affecting our construction. In the region where 
our gauge field configuration has non negligible
field strength the Wilson line can simply be gauged away and 
so can not destroy the fact that our background is a solution to the equations of motion. 
 We can also add other localised  gauge field configurations, located at different positions in 
the internal space, to our background without disturbing our analysis of the moduli of the gauge 
field lump of central interest.

For concreteness, tractability and to keep the number of bundle moduli manageable we shall in 
this section consider a localised gauge field configuration, as described above, which is 
simply a Yang Mills instanton dressed up to fit in the current context. The relevant instanton 
solution to pure Yang-Mills theory is of course well known but we shall 
find that working in terms of the ADHM formalism rather than just writing down a solution is 
necessary if we wish to obtain the K\"ahler potential describing the moduli space of 
the compactification.

\vspace{0.5cm}

Once we have our orbifold and gauge field configuration then our method 
for obtaining the four dimensional effective action associated with heterotic string or M-theory 
compactified on this background is as follows \cite{Gray:2003vw}. One starts with our complete background 
solution and promotes all of the integration constants in the configuration to be four 
dimensional fields. In doing this ambiguities arise in how to promote any field with a four 
dimensional index as it could contain a piece proportional to the four dimensional derivative 
of one of these promoted (moduli) fields. This ambiguity is removed by demanding that any 
solution of the low energy theory we obtain is associated with a 
solution, up to the usual approximations 
such as slowly varying moduli, of the full higher dimensional system. This leads to the concept 
of compensators as discussed in this context in \cite{Gray:2003vw}.

Once we have promoted the integration constants to moduli fields in a consistent way we then 
substitute the resulting configuration into the ten dimensional action and integrate out the 
extra dimensions to leave a four dimensional theory. In doing so it is important to check that 
the result obtained does not depend on 
any field in a region where our approximate vacuum solution, 
as described above, is not valid. For example the result should not depend on the gauge field 
far from the centre of any of our localised lumps.

When we follow this procedure, assuming we take care of a few subtleties such as making sure 
we use a sensible definition for various four dimensional moduli, we obtain a four dimensional 
action which contains kinetic terms for both metric and gauge bundle moduli. The moduli space 
metric associated with these kinetic terms is of course K\"ahler as the model is $N=1$ 
supersymmetric by construction and these scalar fields are components of four dimensional 
chiral superfields. We can therefore write down a K\"ahler potential that encodes our result and 
in fact the ability to do so provides a nice check that we have got the, reasonably lengthy, 
calculation correct.

The details of this kind of calculation are described at length in \cite{Gray:2003vw} and so 
we shall not reproduce them here. Rather we shall, later on in this section, take some results 
from that paper as our starting point and refer the interested reader to the original 
for details.

In \cite{Gray:2003vw} only a few of the moduli of the instanton under consideration where kept, namely 
a modulus describing its size (in the transverse four dimensional space) and 
the orientation of the gauge field expectation value 
within the $SU(2)$ subgroup of the full gauge group of the 
heterotic model in which it is embedded. 
Here we wish to include some extra moduli - 
embedding moduli which describe how this $SU(2)$ subgroup is embedded within the overall gauge 
group of the model. 

In introducing these new fields we encounter a problem. We are going to end up with a much more 
complicated four dimensional theory than that obtained in earlier work 
and the method used previously to 
obtain the K\"ahler potential from the component action, essentially trial and error, 
is no longer going to be 
viable.

Therefore we shall begin by studying a Yang-Mills instanton in pure gauge theory using
 the ADHM formalism and seeing what the K\"ahler potential associated with the 
moduli space of this object looks like. We will then reintroduce the rest of our fields and will
find that, using the information we have gained, we can argue for a form for the complete 
K\"ahler potential. Once this has been obtained it can easily be shown to be correct by 
direct calculation of the component action.

\subsection{The ADHM formalism and the `real' and `holomorphic' descriptions of hyperk\"ahler 
quotients.}
\label{adhm}

Let us consider the moduli spaces of solutions garnered from the ADHM construction 
\cite{Dorey:2002ik}. In 
particular we shall specialise to consider a single $SU(2)$ instanton embedded within an overall
$SU(3)$ living in four dimensional flat space \cite{Bernard:1979qt}. 

In discussing the ADHM construction our notation will be the same as ref \cite{Dorey:2002ik}. The ADHM 
construction provides us with a self-dual gauge field configuration in terms of a certain set 
of constrained data. We shall not provide 
a self contained general 
discussion of the construction here but shall simply describe the elements 
which are directly relevant to the case in hand. Various definitions and conventions pertinent to this 
subsection can be found in appendix A.

The central object of the ADHM construction is a complex-valued matrix 
$\Delta_{(u+\alpha) \dot{\alpha}}$ (where the reader familiar with the formalism 
will note we have already specialised the 
general construction to the single instanton case outlined above). This matrix is linear in the 
(four dimensional) spatial coordinates.

\begin{eqnarray}
\Delta_{(u+\alpha) \dot{\alpha}}(x) = a_{(u+\alpha) \dot{\alpha}} + b^{\beta}_{(u+\alpha)} 
x_{\beta \dot{\alpha}}
\end{eqnarray}
Here $x_{\alpha \dot{\alpha}}$ are the four dimensional spatial 
coordinates in a quaternionic basis, 
$x_{\alpha \dot{\alpha}} = x_A \sigma_{A \alpha \dot{\alpha}}$ where $A$ is the usual four 
dimensional vector index, $\sigma_{A} = (i \vec{\tau}, 1_{[2]\times[2]})$ and the $\vec{\tau}$ 
are the three Pauli matrices.

The gauge field is then constructed from a set of complex matrices which form a basis of the 
null space of $\Delta$ and its conjugate. We define these matrices, $U$, as follows.

\begin{eqnarray}
\bar{\Delta}^{\dot{\alpha} (u+\alpha)} U_{(u+\alpha) v} = 0 = \bar{U}^{(u+\alpha)}_v 
\Delta_{(u+\alpha) \dot{\alpha}}
\end{eqnarray}
We orthonormalise U.
\begin{eqnarray}
\bar{U}^{(u+\alpha)}_v U_{(u+\alpha) w} = \delta_{vw}
\end{eqnarray}
The self-dual gauge field configuration we are interested in 
is then given in terms of these matrices.
\begin{eqnarray}
(A_{B})_{vw} = i \bar{U}^{(u+\alpha)}_v \partial_B U_{(u+\alpha) w}
\end{eqnarray}

 That such a definition of our gauge field leads to a self-dual field strength can then be easily 
shown (see ref \cite{Dorey:2002ik}) provided that our ADHM data obeys the addition constraint,
\begin{eqnarray}
\label{constraint1}
\bar{\Delta}^{\dot{\alpha} (u+\alpha)} \Delta_{(u+\alpha) \dot{\beta}} = 
\delta^{\dot{\alpha}}_{\dot{\beta}} (f^{-1}).
\end{eqnarray}
In the above $f$ is an arbitrary function.

This last equation is what leads us to the famous ``ADHM constraints''. If we use some of the 
freedom afforded by the redundancy in the description of the gauge field in terms of the ADHM 
data we can put $a$ and $b$ in the following form.

\begin{eqnarray}
b^{\beta}_{(u+\alpha)} = \left( \ba{c} 0\\  \delta_{\alpha}^{\beta} \ea \right) \;\;\; &,& \;\;\;
\bar{b}_{\beta}^{(u+\alpha)} = \left( 0 \;\;\;  \delta^{\alpha}_{\beta}  \right) \\
a_{(u+\alpha) \dot{\alpha}} =  \left( \ba{c} \omega_{u \dot{\alpha}} \\ a'_{\alpha \dot{\alpha}}
 \ea \right) \;\;\; &,& \;\;\;  \bar{a}^{\dot{\alpha} (u + \alpha)} = \left( 
\bar{\omega}^{\dot{\alpha}}_u \;\;\; \bar{a}'^{\dot{\alpha} \alpha}  \right) 
\end{eqnarray}

Henceforth we shall set $a' =0$ as this data corresponds to the position moduli of the instanton.
 Our method is not amenable to describing these moduli within the heterotic context and they
are not of interest in any case in terms of 
our goals as stated in the introduction.
The constraints which follow from equation \eqref{constraint1} can then be reduced to 
the following.
\begin{eqnarray}
\label{adhmconstraint}
\vec{\tau}^{\dot{\alpha}}_{\dot{\beta}} (\bar{\omega}^{\dot{\beta}} \omega_{\dot{\alpha}} ) = 0
\end{eqnarray}
We also find the following expression for the function $f$.
\begin{eqnarray}
f = 2 \left( \bar{\omega}^{\dot{\alpha}} \omega_{\dot{\alpha}} + 2 (x_B)^2  \right)^{-1}
\end{eqnarray}

Despite the choices we have made there is still some redundancy in our data $\omega$ in terms of 
describing our self-dual gauge field configuration. In particular, for our single instanton case
 the physical gauge field is invariant under $U(1)$ transformations of the ADHM data of the 
form,
\begin{eqnarray}
\label{U1}
\omega_{\dot{\alpha}} \rightarrow \omega_{\dot{\alpha}} \Gamma  \;\;  .
\end{eqnarray}
In this expression $\Gamma$ is the $U(1)$ transformation. 

\vspace{0.5cm}

Before going on to describe how the ADHM formalism is an example of a hyperk\"ahler quotient 
construction lets make things a little more explicit by writing out how all of the above looks 
for the special case of our single instanton embedded within $SU(3)$.

A solution to the ADHM constraints is given in this case by,

\begin{eqnarray}
\label{omegadef}
\omega = \rho U_3 U_2 m_1  \;\; .
\end{eqnarray}
Here we have,
\begin{eqnarray}
U_2 &=& \left( \ba{rr} \theta^{\gamma} \sigma_{\gamma} &  \ba{rr} 0 \\ 0 \ea \\ 0 \;\;\; 0 & 0 \;
\ea \right)\\
U_3 &=& \left( \ba{rrr} 
\frac{a_3^2+a_4^2 + (a_1^2+a_2^2)\cos{a}}{a^2} &
\frac{(a_1 -i a_2)(a_3 + i a_4)(\cos{a}-1)}{a^2} & 
\frac{\sin{a} (a_1 i + a_2) }{a} \\ 
\frac{(a_1 + i a_2)(a_3 - i a_4)(\cos{a}-1)}{a^2} &
\frac{a_1^2+a_2^2 + (a_3^2+a_4^2)\cos{a}}{a^2}  & 
\frac{\sin{a} (a_3 i + a_4) }{a}  \\ 
\frac{\sin{a} (a_1 i - a_2) }{a} & 
\frac{\sin{a} (a_3 i - a_4) }{a} & 
\cos{a} \;\;\;\;  \ea \right)
\\
\label{m1def}
m_1 &=& \left(\ba{rrr}1&0\\0&1\\0&0\ea\right) \;\;\; , \textnormal{where} \\
a &=& \sqrt{a_1^2+a_2^2+a_3^2+a_4^2}
\end{eqnarray}

Thus, with the constraint $\sum_{\gamma =1}^4 (\theta^{\gamma})^2 = 1 $, $U_2$ 
is an $SU(2)$ matrix 
which will describe the orientation of our instanton within $SU(2)$. The matrix $U_3$ lies in
 $\frac{SU(3)}{SU(2) \times U(1)}$ and will describe the embedding of our gauge field 
configuration within $SU(3)$. We have chosen to split up the matrix $U_3 U_2$, which takes values in 
$\frac{SU(3)}{U(1)}$ like this in order to separate the associated moduli in a manner which will 
be useful to us in later sections. Specifically, with the description we have here, the $\theta$'s are
orientation moduli for any values of the embedding moduli $a$. As such only the $a$'s describe the 
instanton `moving through the gauge group'. 
Finally we will see when we come to write down the gauge field 
itself that $\rho$ can be interpreted as the size of our instanton in the four dimensional space. 
In writing down these expressions we have made a particular choice of gauge for the $U(1)$ ambiguity in the 
data \eqref{U1}.

From this explicit form of the ADHM data we can of course derive explicit forms of the other 
quantities we shall need. For example we find \cite{Dorey:2002ik},

\begin{eqnarray}
f &=& \frac{1}{x^2+\rho^2} \\
U_{(u+\alpha)v} &=& \left(\ba{rr} 1_{[3] \times [3]} + \frac{1}{\rho^2} \left( 
\sqrt{\frac{x^2}{x^2+\rho^2}} -1 \right) \omega_{\dot{\alpha}} \bar{\omega}^{\dot{\alpha}} \\  
- \left( x_{\alpha \dot{\alpha}} \bar{\omega}^{\dot{\alpha}}\right)/\left(|x| 
\sqrt{x^2 +\rho^2} \right) \,\;
\ea \right) 
\end{eqnarray}

Finally from the matrix basis, $U$, of the null space of the ADHM operator we can write down 
the explicit expression for the gauge field (written here in singular gauge).

\begin{eqnarray}
A_B = 2 i \frac{\omega_{\dot{\alpha}} 
x_C \bar{\sigma}_{CB \,\,\, \dot{\beta}}^{\,\,\,\,\,\,\,\,\dot{\alpha}} 
\bar{\omega}^{\dot{\beta}} }{x^2 \left( x^2 + \rho^2 \right) }
\end{eqnarray}

In this expression $\bar{\sigma}_{CB} = \frac{1}{4}\left(\bar{\sigma}_C \sigma_B - \bar{\sigma}_B 
 \sigma_C \right) $ as usual.
We can now see what the residual $U(1)$ transformation of equation \eqref{U1} corresponds to 
in this case. It is simply the $U(1)$ generated by the
eighth Gell-Man matrix $\lambda_8$, which leaves the instanton configuration unchanged.

\vspace{0.5cm}

Returning to the main discussion, 
we are now in a position to see how the ADHM construction is in fact an example of a 
hyperK\"ahler quotient construction \cite{Hitchin:1986ea}. As the name suggests, a hyperK\"ahler quotient 
construction is a method of creating one hyperK\"ahler manifold from another. In the current 
context the manifold being created is the eight dimensional 
moduli space of the instanton configuration and 
the mother space is simply flat space in twelve dimensions.

The moduli space of the instanton configuration is parameterised by the data $\omega$ and 
$\bar{\omega}$ subject to the constraints \eqref{adhmconstraint} and upon gauging of the 
isometry \eqref{U1}. This isometry is generated by,
\begin{eqnarray}
\label{X}
X =  i \bar{\omega}^{\dot{\alpha}}_u  \frac{\partial}{\partial \bar{\omega}^{\dot{\alpha}}_u} 
- i \omega_{u \dot{\alpha}} \frac{\partial}{\partial \omega_{u \dot{\alpha}}}  \;\; . 
\end{eqnarray}

The form of moduli space outlined in the previous paragraph turns out to be precisely that
 of a hyperK\"ahler quotient of twelve 
dimensional flat space by the isometry \eqref{X}. The three constraints \eqref{adhmconstraint}
 correspond to the restriction to the level set, which is defined by the vanishing of the moment 
maps associated with the isometry with which we wish to quotient (where in this application 
we set the central elements to be zero). 
The gauging of the isometry \eqref{U1} then corresponds to the final step of the construction 
where we take an ordinary quotient of the level set by the isometry we are interested in 
\cite{Hitchin:1986ea}.

The fact that it is the flat metric we are taking the quotient of in the ADHM construction can be 
ascertained by examining the moduli space metric for the instanton in terms of the variables 
$\omega$ and $\bar{\omega}$ \cite{Dorey:2002ik}. 
The metric on the quotient space is then simply that induced 
on the submanifold the construction gives by the flat metric on the mother space.

The moduli space of our instanton configuration is then
given by the construction outline above which I shall call the `real formulation'. However this 
description of the hyperK\"ahler quotient is not the best one from which to recover the complex 
structure associated with the daughter space. This is due to the fact that we obtain the quotient 
by imposing 3 constraints and then removing the last direction by the gauging of a real isometry. 
This structure clearly does not sit ideally with a pairing up of coordinates into complex 
parameters. Thus, although the complex structure does descend in a nice manner from the mother 
space to the daughter space in this formalism, it is considerably easier to obtain a set of 
complex coordinates over the quotient space by going to what we shall 
refer to as the `holomorphic formulation'.

The idea behind this formulation of the hyperK\"ahler quotient (which is described in detail in 
\cite{Hitchin:1986ea}) is that the construction outlined 
above is in fact equivalent to imposing two of the three level set conditions and then taking a 
quotient with respect to the complexification of the isometry \eqref{X}. This form of 
the construction preserves the holomorphic structure of the coordinates throughout and so 
we obtain a set of complex coordinates, and a K\"ahler potential, on our moduli space 
automatically.

Let us choose a set of holomorphic coordinates to be $z= \left( \bar{\omega}^{\dot{2}}_u , 
\omega_{\dot{1} u}\right)$.
The holomorphic/anti-holomorphic part of the level set constraints can be written as follows 
\cite{Hitchin:1986ea}.
\begin{eqnarray}
\label{hollevel}
\bar{\omega}^{\dot{2}}_u \omega_{\dot{1} u} = 0 \;\;\; , \;\;\; \bar{\omega}^{\dot{1}}_u 
\omega_{\dot{2} u} = 0
\end{eqnarray}
These constraints are preserved by the complexification of the isometry which is generated by 
\cite{Hitchin:1986ea},
\begin{eqnarray}
\label{compisom}
i \bar{\omega}^{\dot{2}}_u \frac{\partial}{ \partial \bar{\omega}^{\dot{2} u}} - i 
\omega_{u \dot{1}} \frac{\partial}{\partial \omega_{u \dot{1}}}  \;\;\; , \;\;\; 
i \bar{\omega}^{\dot{1}}_u \frac{\partial}{ \partial \bar{\omega}^{\dot{1} u}} - i 
\omega_{u \dot{2}} \frac{\partial}{\partial \omega_{u \dot{2}}} \;\;\; .
\end{eqnarray}

Thus to obtain the K\"ahler potential and moduli space of our instanton all we need to do is 
the following. We start with the K\"ahler potential for twelve dimensional flat space,
\begin{eqnarray}
\bar{\omega}^{\dot{\alpha}}_u \omega_{u \dot{\alpha}} \;\;\;  .
\end{eqnarray}
We then divide out by the complexified isometry \eqref{compisom} by introducing an 
auxiliary vector 
superfield, $V$, which gauges the isometry \cite{Hull:1985pq}.
\begin{eqnarray}
K= e^{V} \bar{\omega}^{\dot{2}}_{u} \omega_{\dot{2} u} +  
e^{-V} \bar{\omega}^{\dot{1}}_{u} \omega_{\dot{1} u} 
\end{eqnarray}

If we define $V$ such that it changes under the isometry in an appropriate manner, 
then the resulting K\"ahler 
potential is invariant under the transformation associated with \eqref{compisom}. We find, 
\begin{eqnarray}
\omega_{\dot{1}} &\rightarrow& e^{\bar{\Lambda}(\bar{z})} \omega_{\dot{1}} 
\;\;\; \bar{\omega}^{\dot{1}} 
\rightarrow e^{-\Lambda(z)} \bar{\omega}^{\dot{1}}\\
\omega_{\dot{2}} &\rightarrow& e^{\Lambda(z)} \omega_{\dot{2}} \;\;\; \bar{\omega}^{\dot{2}} 
\rightarrow e^{-\bar{\Lambda}(\bar{z})} \bar{\omega}^{\dot{2}}\\
V &\rightarrow& V + \bar{\Lambda}( \bar{z}) - \Lambda (z)
\end{eqnarray}
We can use this to transform to 
a new set of coordinates, $\left( \bar{\omega}' \;, \; \omega' \right)$, where 
$\bar{\omega}'^{\dot{2}}_1$
, for example, is simply $1$. Thus we have indeed removed one of the complex dimensions by 
quotienting by the complex isometry.

Next, since $V$ is not a true dynamical field but merely auxiliary we can integrate it out by using its 
equation of motion to obtain, 

\begin{eqnarray}
K= 2 \sqrt{\bar{\omega}'^{\dot{2}}_{u} \omega'_{\dot{2} u} 
\bar{\omega}'^{\dot{1}}_{v} \omega'_{\dot{1} v}} \;\;\; .
\end{eqnarray}

Finally, since the holomorphic part of the level set constraint is invariant under 
the complexified isometry we can simply impose \eqref{hollevel} 
at this stage to eliminate one more 
complex coordinate and obtain the
complex structure and K\"ahler potential for the eight dimensional manifold that 
we desire.

\begin{eqnarray}
\label{adhmk}
K=2 \sqrt{(1+ |C_1|^2 + |C_2|^2)(|C_1 C_3 + C_2 C_4|^2 + |C_3|^2 + |C_4|^2)}
\end{eqnarray}

\begin{eqnarray}
C_1 &=&  \frac{\bar{\omega}^{\dot{2}}_{2}}{\bar{\omega}^{\dot{2}}_{1}} \;\;\;\;
C_2 =  \frac{\bar{\omega}^{\dot{2}}_{3}}{\bar{\omega}^{\dot{2}}_{1}} \\
C_3 &=&  \bar{\omega}^{\dot{2}}_{1} \omega_{\dot{1} 2} \;\;\;\;
C_4 = \bar{\omega}^{\dot{2}}_{1} \omega_{\dot{1} 3} 
\end{eqnarray}
The complex coordinates can be written in terms of the component fields 
$\left( \rho, \theta^{\gamma}, a^{I} \right)$ 
which describe the size, $SU(2)$ orientation and embedding 
of the instanton within the gauge group respectively using equations \eqref{omegadef} - 
\eqref{m1def}. The complex coordinates given here 
are a sensible choice when $\bar{\omega}^{\dot{2}}_1 \neq 0$. 
Obviously we may make other, very similar, choices of coordinates to cover patches of the moduli space 
where this inequality does not hold.

Further mathematical details of the structure of the space we have been dealing with here can be found in 
\cite{Vandoren:2000qr}. 
It is a cone over a 3-Sasakian space which is itself a non-trivial fibration of $Sp(1)$ over a 
Wolf space.

We shall shortly  see that the K\"ahler potential we have presented in this subsection is what 
we wish to embed within the M-theory context.


\subsection{Including the gravitational moduli.}
\label{grav}

We shall now proceed to obtain the component form of the four dimensional effective action 
describing the embedding and other instanton moduli within our M-theoretic context. This 
component form of the action will be matched with a K\"ahler potential, obtained using the 
discussion of the previous subsection as a guide, in the next subsection.

Our starting point is the ten dimensional effective action of heterotic models. The relevant pieces for our 
purposes are,

\begin{eqnarray}
\label{10Daction2}
S_{10} = \frac{1}{2 \k^2_{10D}} \int d^{10}x \sqrt{-g} \;  e^{2\phi} 
\left( -R - 4(\partial \phi)^2 + 
\frac{1}{3} H^2 + 2 \a' \textnormal{tr} F^2  +... \right) \; .
\end{eqnarray}
Here the three-form field strength, $H$, is defined (at least locally) as follows.
\begin{eqnarray}
\label{10DHdef}
H= dB - 2 \a' \omega_{3YM} + ...
\end{eqnarray}
The Chern-Simons three form associated with the gauge field has been denoted as $\omega_{3YM}$. 
The traces in these expressions, and those in the rest of the paper, 
are in the fundamental of $SU(3)$. 

We are going to dimensionally reduce this action on the background solution described in subsection 
\ref{method} (which will be given in more mathematical detail below) \cite{Gray:2003vw}. 
When we dimensionally reduce we shall keep various integration constants of the higher dimensional 
solution in the form of four dimensional moduli fields. We will not keep all of the constants which would 
correspond to embedding moduli however. In fact we shall consider only those 
embedding moduli which describe the instanton moving in 
an $SU(3)$ subspace of the entire gauge group. We could
work with the instanton embedded in the complete gauge group of the $SO(32)$ heterotic string, 
for example, \cite{Dorey:2002ik}; in other words 
we could keep all of the embedding moduli. Our method would be essentially unchanged from the one 
described in this paper. We will stick, however, to this 
simpler case so that we can keep the number of moduli manageable thus making the explicit
computation of the component action as described in this subsection 
 practical. To give an idea of the simplification this affords us consider the following. The 
little group of $SU(2)$ 
within $SU(3)$ is $U(1)$ which is one dimensional. Thus, since $SU(3)$ is eight dimensional and 
we have three orientation moduli, the number of embedding moduli we shall have to deal with 
is $8 - 3 - 1 =4$. In contrast 
if we wanted to keep all of the embedding moduli and work with, say, the minimal embedding in the
 $SO(32)$ case the relevant little group would be $SU(2) \times 
SO(28)$ which has $381$ dimensions. Thus we would end up with $496 - 3 - 381= 112$ 
embedding moduli in this example. 

Given this truncation and after 
having promoted the integration constants of the higher dimensional background 
solution to be four dimensional fields 
in preparation for the reduction the relevant portions of the 
vacuum configuration appears as follows \cite{Gray:2003vw}.

\begin{eqnarray}
\label{bgsoln}
A_B &=& U_3 
\sigma_{\gamma} \theta^{\gamma} \frac{2 i \rho^2 r^C \bar{\sigma}_{CB} {\cal V}_{(B)}^{\frac{1}{6}}
 }{R^2(R^2+ \rho^2)} 
\bar{\sigma}_{\delta} \theta^{\delta} \bar{U}_3 \\ 
A_{\mu} &=& \Omega^{(A)}_{{\it m}} \partial_{\mu} {\it m} \\ \label{phi0def}
e^{-2 \phi} &=& e^{-2 \phi_0} \left( 1 + 8 \a' \frac{R^2 + 2 \rho^2}{(R^2 + \rho^2)^2} \right) \\ 
\label{themetric}
ds^2_{10} &=& h g_{\mu \nu} dx^{\mu} dx^{\nu} + \cV_3^{\frac{1}{3}} \delta_{ MN }dx^M dx^N +  
e^{2 \phi_0 - 2 \phi} (\cV_{(A)}^{\frac{1}{3}} \delta_{AB} dx^A dx^B) \\ \nonumber
&& + 2 \Omega^{(g)}_{(a |{\it m}|} \partial_{\mu)} {\it m} dx^a dx^{\mu}  \\
B_{AB} &=& B^{bg}_{AB} + \frac{1}{6} \chi_{(A)} \Pi^{(A)}_{AB} \\
B_{MN} &=& B^{bg}_{MN} + \frac{1}{6} \chi_{(3)} \Pi^{(3)}_{MN}  \\
B_{a \mu} &=& \Omega^{(B)}_{[a |{\it m}|} \partial_{\mu ]} {\it m} \\ \label{bggaug2}
B_{\mu \nu} &=& \textnormal{constant}
\end{eqnarray}

There are many quantities in the above configuration which we need to define. Firstly, $A,B=6..9$ 
are indices associated with directions transverse to the instanton in our hidden space, $M,N =4,5$
 are associated with the other two dimensions in the Calabi-Yau and $\mu,\nu = 0..3$ are the 
indices associated with four dimensional space and time.
Secondly, ${\cal V}_{(1)} = {\cal V}_{(2)} = {\cal V}_{1}$, and 
${\cal V}_{(3)} = {\cal V}_{(4)} = {\cal V}_{2}$ are four dimensional fields describing the 
size of the transverse space to the instanton within the Calabi-Yau (The reason for this slightly 
unusual notation is described in \cite{Gray:2003vw}). Similarly 
${\cal V}_{(5)} = {\cal V}_{(6)} = {\cal V}_{3}$ describes the size of the two cycle in the 
compact space which our gauge field configuration is extended in. We choose the function
of four dimensional fields $h$ so that the reduction gives us
a canonically normalised Einstein Hilbert term. We find $h = {\cal V}_1^{-\frac{1}{3}} 
{\cal V}_2^{-\frac{1}{3}} {\cal V}_3^{-\frac{1}{3}} e^{- 2 \phi_0}$. The four dimensional field 
$\phi_0$ is related to the four dimensional dilaton. We define $(r^C) = (\cV_1^{\frac{1}{6}} x^6,
 \cV_1^{\frac{1}{6}} x^7 , \cV_2^{\frac{1}{6}} x^8 , \cV_2^{\frac{1}{6}} x^9)$ and 
$R^2 = \delta_{AB} r^A r^B$. The background two-form is denoted $B^{bg}$ and the 
fields $\chi$ will become four dimensional axions. We write arbitrary moduli as $m$. 
We have denoted three 
harmonic two forms on a flat orbifold of our type as (in the absence of the instanton)
 $\Pi^{(1)}_{ab} = (+1|_{a=6,b=7},-1|_{a=6,b=7})$,
 $\Pi^{(2)}_{ab} = (+1|_{a=8,b=9},-1|_{a=8,b=9})$ and  
$\Pi^{(3)}_{ab} = (+1|_{a=4,b=5},-1|_{a=4,b=5})$.

This just leaves the objects $\Omega$. These are the compensators 
which were discussed earlier - contributions to the reduction ansatz which are necessary if a 
solution to the lower dimensional action we are to obtain is to be associated with an appropriate 
approximate solution to the higher dimensional system. As was explained in detail in \cite{Gray:2003vw} 
the compensators associated with the metric and two form fields, $\Omega^{(g)}$ and 
$\Omega^{(B)}$ do not enter into the calculation of the four dimensional effective theory and 
so we shall not need their specific form. The gauge field compensators, $\Omega^{(A)}$, do 
play an important role in the dimensional reduction however. In the case where the embedding 
moduli, $a^I$, are set to zero these compensators were written down explicitly in \cite{Gray:2003vw}. To 
find the compensators for the case at hand we merely need to take those expressions and apply 
the global gauge transformation associated with $U_3$.

\begin{eqnarray}
\label{gc}
\Omega^{(A)}_m = U_3 \Omega^{(A)}_{m \; (a=0)} \bar{U}_3
\end{eqnarray}

We will not therefore reproduce all of these expressions again explicitly here. The reader who 
requires the precise form of these quantities can consult reference \cite{Gray:2003vw}.

This then just leaves the compensators $\Omega^{(A)}_m$ where $m=a^I$. Expressions for these 
compensators in the absence of the metric moduli are furnished by the ADHM formalism \cite{Dorey:2002ik}.

\begin{eqnarray}
\label{omegaa}
\nonumber
\Omega^{(A)}_{a^{I}} = -i \left\{ \left(1- \sqrt{\frac{x^2}{x^2+\rho^2}} \right) 
\left( \frac{\partial U}{\partial a^{I}} \left(\ba{rrr} 1&0&0 \\ 0&1&0 \\ 0&0&0 \ea \right)
\bar{U} -  U \left(\ba{rrr} 1&0&0 \\ 0&1&0 \\ 0&0&0 \ea \right)
\frac{\partial \bar{U}}{\partial a^{I}} \right) + \right. \\ 
U \left(\ba{rrr} 1&0&0 \\ 0&1&0 \\ 0&0&0 \ea \right) \frac{\partial \bar{U}}{\partial a^I} U 
\left(\ba{rrr} 1&0&0 \\ 0&1&0 \\ 0&0&0 \ea \right) \bar{U} \left(1- \sqrt{\frac{x^2}{x^2+\rho^2}}
 \right)^2   \\  \nonumber \left.
- \frac{\rho^2}{x^2+\rho^2} U 
\left(\ba{rrr} 1&0&0 \\ 0&1&0 \\ 0&0&0 \ea \right) \bar{U} \frac{1}{4} {\textnormal tr}\left[ 
\left(\ba{rrr} 1&0&0 \\ 0&1&0 \\ 0&0&0 \ea \right) \left( \bar{U} \frac{\partial U}{\partial a^I}
 - \frac{\partial \bar{U}}{\partial a^I} U \right)\right] \right\}
\end{eqnarray}

We need only then reintroduce the metric moduli via coordinate transformations - in other 
words by replacing $x^2$ in the above by $R^2$. As before this procedure is 
described in detail in ref \cite{Gray:2003vw}.

We are now in a position to begin the dimensional reduction. As was explained in \cite{Gray:2003vw},
 it turns out that not every term in the action \eqref{10Daction2} 
contributes to the terms in the 
four dimensional 
effective action which involve bundle moduli. In fact the only two terms which do contribute to 
order $\a'$ are 
the following ones.
\begin{eqnarray}
\label{2terms}
\frac{1}{2 \k^2} \int d^{10} x \sqrt{-g} \; e^{2 \phi} \left( - \frac{4}{3} \a' \textnormal{d} 
B \omega_{3 YM} + 2 
\a' \textnormal{tr} F^2 \right)\; .
\end{eqnarray}

There are various subtleties involved in the calculation that brings us to the conclusion that 
these are the only terms which contribute. Firstly one has to use a small amount of knowledge 
that we have about the full {\it exact} solution describing the Calabi-Yau and gauge bundle. 
For instance it is necessary to use the fact that the Calabi-Yau is compact and a small amount 
of information about the index structure of the compensators. Another subtlety is that one has 
to be careful that the definition of the four dimensional fields in the solution which are 
to become moduli are sensible ones. This involves shifting the definition of $\phi_0$, 
for example, by a constant from the one presented in \eqref{phi0def}. Detailed discussions of 
these issues are presented in \cite{Gray:2003vw}.

One could proceed to substitute equations \eqref{bgsoln}-\eqref{omegaa} into \eqref{2terms} 
and integrate out the 
extra dimensions at this point. However it turns out to be calculationally more expedient to 
write the second term in terms of zero modes (defined below) first. One may then obtain the 
zero modes for our new fields, the $a$ moduli, directly from the ADHM formalism with a considerable
saving in time and effort as compared to a direct calculation of these quantities from the gauge 
field and compensators.

Rewriting equation \eqref{2terms} we find,
\begin{eqnarray}
\label{2terms2}
\frac{1}{2 \k^2} \int d^{10} x \sqrt{-g}\; e^{2\phi} \left(-\frac{4}{3} \a' 
{\cal V}_{(A)}^{-\frac{2}{3}}\partial^{\mu} \chi^{(A)} \Pi^{(A)}_{AB} \omega^{3YM}_{AB \mu}  
  + 4 \a' {\textnormal tr}\left\{ 
{\cal V}_{(A)}^{-\frac{1}{3}} Z_{A \; m} Z_{A \; n}  \right\} \partial_{\mu} m \partial^{\mu} n 
\right) \; .
\end{eqnarray}
Here the Chern-Simons three form, $\omega^{3YM}_{\mu A B}$, 
is built out of the background gauge field in the usual manner and so is 
proportional to the derivative of a modulus field. The four dimensional indices have been raised using 
the four dimensional metric.

The zero modes are defined by, 

\begin{eqnarray}
F_{A \mu} = Z_{A m} \partial_{\mu} m \;\;\;.
\end{eqnarray}
These zero modes are unchanged from reference \cite{Gray:2003vw} for $m \neq a^I$ except for a global 
gauge transformation of the same form as in 
\eqref{gc}. As mentioned above, the zero modes with $a$ indices are supplied to us,
in the absence of metric moduli, by the ADHM construction \cite{Dorey:2002ik}. 
\begin{eqnarray}
\label{azeromode}
\nonumber
Z_{A a^I} = \frac{i \rho^2}{x (x^2 + \rho^2 )^{\frac{3}{2}}} \left[U U_2 
\left(\ba{rrr} 1&0 \\ 0&1 \\ 0&0  \ea \right) x^B \bar{\sigma}_B \sigma_A 
\left(\ba{ccc} 1&0&0 \\ 0&1&0 \ea \right) \bar{U}_2 \left( \frac{\partial \bar{U}}{\partial a^I} 
- \frac{1}{4} \bar{U} {\textnormal tr} \left[ \left(\ba{rrr} 1&0&0 \\ 0&1&0 \\ 0&0&0 \ea \right) 
\left( \bar{U} \frac{\partial U}{\partial a^I}
 - \frac{\partial \bar{U}}{\partial a^I} U \right) \right] \right)  \right. \\ 
 \left(1 + U \left(\ba{rrr} 1&0&0 \\ 0&1&0 \\ 0&0&0 \ea \right) \bar{U} 
 \left( \sqrt{ \frac{x^2}{x^2+ \rho^2}} -1\right)  \right) \\ \nonumber  - 
\left(1+ U \left(\ba{rrr} 1&0&0 \\ 0&1&0 \\ 0&0&0 \ea \right) \bar{U} 
\left(\sqrt{\frac{x^2}{x^2+ \rho^2}} -1\right)  \right)
 \left( \frac{\partial U}{\partial a^I} 
+ \frac{1}{4} U {\textnormal tr}\left[\left(\ba{rrr} 1&0&0 \\ 0&1&0 \\ 0&0&0 \ea \right) 
\left( \bar{U} \frac{\partial U}{\partial a^I}
 - \frac{\partial \bar{U}}{\partial a^I} U \right) \right] \right) \\ \nonumber \left.
 U_2 \left(\ba{rrr} 1&0 \\ 0&1 \\ 0&0  \ea \right) 
\bar{\sigma}_A x^B \sigma_B \left(\ba{ccc} 1&0&0 \\ 0&1&0 \ea \right) 
\bar{U}_2 \bar{U} \right]
\end{eqnarray}
We can then reintroduce the metric moduli by coordinate transformations in exactly the same way 
we did for the $a^I$ field gauge compensators.

Finally we can simply put our various expressions together and perform the integrations over the 
hidden dimensions to obtain the four dimensional effective action in component form. We shall 
not however present the result in this form here. This is because the resulting moduli space metric
is extremely complex, being a highly non-trivial structure over 16 dimensions, and the relevant expression 
would significantly add to the length of the paper without adding much which is useful to 
the reader. Instead we shall present this result in a much cleaner form in the next section where
we shall, using the results of subsection \ref{adhm}, 
match it to a K\"ahler potential and complex structure. 

\subsection{The full result.}
\label{full}

We will now find the K\"ahler potential and complex structure which reproduce the 
component action whose derivation was outline in the previous section. In doing so we will be 
guided by a number of pieces of information which we already have about these quantities.

\begin{itemize}
\item We know what the K\"ahler potential and complex structure look like in the limit as $
\a' \rightarrow 0$. This is simply the normal result for ten dimensional heterotic models 
compactified in the way considered here with the bundle moduli ignored.

\item We also know what these quantities look like if we simply `turn off' (i.e. set constant) 
all of the metric moduli. In this case we should simply be left with the equivalent structures 
for a pure Yang Mills instanton embedded within $SU(3)$. 
This was what we obtained in subsection \ref{adhm}.

\item We know what the K\"ahler potential and complex structure look like in the limit where we 
set all of the embedding moduli, $a_I$, to be constant. This was the result obtained in reference 
\cite{Gray:2003vw}. The fact that this result should be unchanged for constant but non zero $a_I$ (
the results 
of \cite{Gray:2003vw} 
cover the case $a_I=0$) can easily be deduced by examining \eqref{2terms2} and how the 
zero modes, compensators and gauge fields change under the relevant global phase transformation. 

\end{itemize}

So what are we going to write down 
for our K\"ahler potential and complex structure? Clearly we shall 
require the standard zeroth order in $\a'$ result in the absence of bundle moduli together with 
order $\a'$ corrections to this depending on the new fields. From the second point above we see 
that the order $\a'$ correction we need to add to $K$ is simply going to be the K\"ahler 
potential for the pure Yang-Mills instanton, as presented in \eqref{adhmk}, 
with various factors of 
the metric moduli fields added in. To see what these factors may be we look at the third point 
on our list. It can be seen after some thought that the 
only likely looking structure to try is the following (the corrections to the T superfields 
also follow as a natural guess from the third point on our list).

\begin{eqnarray}
\label{Kfull}
K= &-&\ln{(S+ \bar{S})} - \ln{(T_1+\bar{T_1})}   - \ln{(T_2+\bar{T_2})} - \ln{(T_3+\bar{T_3})}
\\ \nonumber &+& \frac{16 \;q_{G5}}{\sqrt{(T_1+\bar{T_1})(T_2+\bar{T_2})}} 
\sqrt{(1+ |C_1|^2 + |C_2|^2)(|C_1 C_3 + C_2 C_4|^2 + |C_3|^2 + |C_4|^2)}
\end{eqnarray}

\begin{eqnarray}
C_1 &=&  \frac{\bar{\omega}^{\dot{2}}_{2}}{\bar{\omega}^{\dot{2}}_{1}} \;\;\;\;
C_2 =  \frac{\bar{\omega}^{\dot{2}}_{3}}{\bar{\omega}^{\dot{2}}_{1}} \\
C_3 &=& e^{-\frac{\beta_1}{2} - \frac{\beta_2}{2}} \bar{\omega}^{\dot{2}}_{1} 
\omega_{\dot{1} 2} \;\;\;\;
C_4 = e^{-\frac{\beta_1}{2} - \frac{\beta_2}{2}}\bar{\omega}^{\dot{2}}_{1} \omega_{\dot{1} 3} \\
T_1 &=& e^{\beta_1} + \frac{2}{3} i \chi_1 + 4 \a' e^{\frac{\beta_1}{2} - \frac{\beta_2}{2}} 
\sqrt{(1+ |C_1|^2 + |C_2|^2)(|C_1 C_3 + C_2 C_4|^2 + |C_3|^2 + |C_4|^2)} \\
T_2 &=& e^{\beta_2} + \frac{2}{3} i \chi_2 + 4 \a' e^{\frac{\beta_2}{2} - \frac{\beta_1}{2}} 
\sqrt{(1+ |C_1|^2 + |C_2|^2)(|C_1 C_3 + C_2 C_4|^2 + |C_3|^2 + |C_4|^2)} \\ \label{T3}
T_3 &=& e^{\beta_3} + \frac{2}{3} i \chi_3 \;\;\;\; S= e^{\varphi} +\sqrt{2} i \sigma
\end{eqnarray}

Here we have defined $e^{\beta_i} = {\cal V}^{\frac{1}{3}}_{(i)}$ and
$\varphi = 2 \phi_0 + \beta_1 + \beta_2 +\beta_3$ where $\phi_0$ in this expression has been redefined by 
a constant shift from that used in \eqref{phi0def} (see \cite{Gray:2003vw} for more details). We have 
also defined $q_{G5}=\frac{\a' (2 \pi)^2}{V_{\textnormal{trans}}}$ where $V_{\textnormal{trans}}$ is 
the coordinate volume of the space transverse to the instanton. It should be 
remembered that $\omega$ is given in terms of the various moduli scalar fields as follows.
\begin{eqnarray}
\label{omegadefagain}
\omega = \rho  \left( \ba{rrr} 
\frac{a_3^2+a_4^2 + (a_1^2+a_2^2)\cos{a}}{a^2} &
\frac{(a_1 -i a_2)(a_3 + i a_4)(\cos{a}-1)}{a^2} & 
\frac{\sin{a} (a_1 i + a_2) }{a} \\ 
\frac{(a_1 + i a_2)(a_3 - i a_4)(\cos{a}-1)}{a^2} &
\frac{a_1^2+a_2^2 + (a_3^2+a_4^2)\cos{a}}{a^2}  & 
\frac{\sin{a} (a_3 i + a_4) }{a}  \\ 
\frac{\sin{a} (a_1 i - a_2) }{a} & 
\frac{\sin{a} (a_3 i - a_4) }{a} & 
\cos{a} \;\;\;\;  \ea \right)
\left( \ba{rr} \theta^{\gamma} \sigma_{\gamma} &  \ba{rr} 0 \\ 0 \ea \\ 0 \;\;\; 0 
&  0 \; \ea \right) 
\left(\ba{rrr}1&0\\0&1\\0&0\ea\right)
\end{eqnarray}

The important point here is that once we have been guided to write down a K\"ahler potential and 
complex structure such as that given above it is easy to then check if it is in fact correct. 
All we need to do is calculate from it, using the usual formulae, the associated component action 
and compare the result with what we obtained in the previous subsection. When we do this we find 
that the above result precisely reproduces the correct component action and so is indeed the 
correct K\"ahler potential and complex structure.

\vspace{0.5cm}

It is worth making a few general comments about this result before going on to examine its application 
to moduli driven phase transitions.

As mentioned above, it reduces to the known result given in reference \cite{Gray:2003vw} when the embedding 
moduli are constant. This can most 
easily be seen by making two observations. Firstly the K\"ahler potential takes the same form in both 
the previously obtained result and in the above when we express it in terms of real component fields. 
Secondly, for constant $a_I$, the fields $C_i$ are holomorphic functions of the associated 
complex superfields of 
reference \cite{Gray:2003vw}. 
These two facts are sufficient to demonstrate that the result specialises in this manner 
correctly. This result means that all of the comments that were made about the K\"ahler potential given 
in \cite{Gray:2003vw} also apply to this, more general, result.

Another observation that we would like to make is that the K\"ahler potential takes a rather 
special form. It can be shown from equations \eqref{themetric} and equations 
\eqref{Kfull}-\eqref{omegadefagain} that, as it does in 
the absence of bundle moduli, the K\"ahler potential takes the form $K \propto \log{V_{CY_3}}$. 
In other 
words the K\"ahler potential is given by the logarithm of the volume of the compactification manifold. 
In showing this result great care must be taken with the definition of $\phi_0$ in the above formalism. In 
particular it must be remembered that the field $\phi_0$ which appears in the definition of four 
dimensional fields above differs from that in equation \eqref{themetric} by a constant shift (the reader 
who requires more detail on this point is refered to
\cite{Gray:2003vw}). It would 
certainly be of great interest to know whether or not this result is true in general when the bundle 
moduli are included in compactifcations upon Calabi-Yau threefolds 
\footnote{The author would like to thank Andr\'{e} Lukas for very useful discussions on this point and 
for suggesting that it may be true in the first place.}.

It is easy to see how the 
generalisation of the result presented here to include further embedding moduli would proceed. The 
arguments presented in this subsection for the form of the full K\"ahler potential would be essentially 
unchanged. The subsequent confirmation of the result by a comparison to the component action, which 
would be computed in exactly the same manner as described in the previous subsection, would be 
quite tedious however.

\vspace{0.5cm}

The above result is clearly quite complicated. It involves $16$ moduli fields in a moduli 
space metric with highly non-trivial structure. In order to achieve our goal of finding an 
explicit description of a phase transition driven by an embedding modulus we are going to have 
to find a consistent way of simplifying the theory we have to deal with. In other words we need 
to find a consistent truncation of the component action which follows from \eqref{Kfull}-
\eqref{T3} which is 
simple enough to find cosmological solutions of and yet which retains the features we require.

We find that the following theory is a consistent truncation of what has been presented above 
and that it has the properties we desire.

\begin{eqnarray}
S_{4D} = \frac{1}{2 \k_{4D}^2} \int d^4 x \sqrt{-g} \left\{R + \frac{1}{2}(\partial \varphi)^2 
+ \frac{1}{2}( \partial \beta_3)^2 + (\partial \beta)^2 + q_{G5} \left[8 e^{-\beta} (\partial 
\hat{\rho})^2 + 4 e^{-\beta} \hat{\rho}^2 (\partial a_4)^2  \right] \right\}
\end{eqnarray}

Here we have set $\beta_1=\beta_2 = \beta$ and have defined $\hat{\rho} = e^{-\frac{\beta}{2}} 
\rho$. 
Clearly we have also consistently truncated a lot of fields (by consistent truncation here we mean that 
any solution of this simple action will also be a solution of the full theory given above).

We shall recapitulate in detail 
what this action describes at the start of the next section which will deal 
with the analysis of a gauge symmetry breaking phase transition driven by the embedding modulus 
$a_4$. 


\section{Embedding moduli and phase transitions.}
\label{cosm}

The starting point for the study of our moduli driven phase transition is the four dimensional 
effective action given below.

\begin{eqnarray}
\label{s4dtrunc}
S_{4D} = \frac{1}{2 \k_{4D}^2} \int d^4 x \sqrt{-g} \left\{R + \frac{1}{2}(\partial \varphi)^2 
+ \frac{1}{2}( \partial \beta_3)^2 + (\partial \beta)^2 + q_{G5} \left[8 e^{-\beta} (\partial 
\hat{\rho})^2 + 4 e^{-\beta} \hat{\rho}^2 (\partial a_4)^2  \right] \right\}
\end{eqnarray}

In order that this section can be read independently of the more technical considerations that 
proceed it we shall now recapitulate the meaning of the different fields that are present. 
First of all we have the usual fields of four dimensional heterotic models, 
the four dimensional dilaton, $\varphi$, and sizes for the various extra dimensions, $\beta$ and 
$\beta_3$. The size of the extra dimensions in which the Yang-Mills instanton described in the 
introduction is localised are determined by the modulus $\beta$. The size of the remaining 
internal 
directions, in which our gauge field configuration is extended, are described by the modulus $\beta_3$. 
The width of the instanton, relative to the size of the hidden space 
transverse to it, is determined by the modulus $\hat{\rho}$. Finally $a_4$ is a four dimensional 
field which describes how the embedding within the larger 
gauge group of the $SU(2)$ valued gauge background
 changes with four dimensional space and time. It is this final field which will play 
the key role in the following discussion.

\subsection{Cosmological evolution.}

To find cosmological solutions to this action we follow the usual procedure of making a FRW 
ansatz.
\begin{eqnarray}
\label{frwansatz}
ds^2 = -dt^2 + e^{2 \a} d \vec{x}^2
\end{eqnarray}
We then allow all of the four dimensional moduli fields to be functions of time only so as to be 
consistent with the symmetries of this metric configuration.

The equations of motion that result from the action \eqref{s4dtrunc} and the ansatz 
\eqref{frwansatz} are not an example of a Toda system as they are in some related examples 
\cite{Lukas:1996iq,Copeland:2001zp,Gray:2003vk}. 
However it turns out the resulting system of equations is still exactly integrable.
The solutions are given below.

Firstly for the metric moduli, scale factor and four dimensional dilaton we find the following.

\begin{eqnarray}
\label{alpha}
\alpha  &=& \frac{1}{3} \log{\left| \frac{t-t_0}{T} \right|} + \alpha_0 \\
\varphi  &=& p_{\varphi} \log{\left| \frac{t-t_0}{T} \right|} + \varphi_0 \\
\beta_3  &=& p_{\beta_3} \log{\left| \frac{t-t_0}{T} \right|} + \beta_3^0 \\
\label{beta}
\beta &=& p_{\beta} \log{\left| \frac{t-t_0}{T} \right|} - 2 
\log{\left( \left| \frac{t-t_0}{T} \right|^{p_{\beta}} +1  \right)} + \beta_0
\end{eqnarray}

In fact it turns out that this is the precise form of the solution however many of the 
bundle moduli described in the previous section 
we decide to include in our analysis, not just for the consistently truncated theory presented 
in \eqref{s4dtrunc}. The only requirement which needs to be imposed for this to be true is that the 
two sizes associated with the space transverse to the instanton must be set equal, $\beta_1=\beta_2=\beta$.
 This `universal' behaviour 
is essentially due to all of the bundle moduli having the same form of coupling to the metric 
moduli, i.e. all of their kinetic terms have a factor of $e^{-\beta}$.
This observation receives corroboration from the result that equations \eqref{alpha}- \eqref{beta} 
agree with the solutions for 
the same fields in reference \cite{Gray:2003vk}.

Given the above we then find the following behaviour for the size of our instanton
 and for its embedding modulus.
\begin{eqnarray}
\hat{\rho} &=& \frac{\left( \hat{\rho}^2_1 \left(\left| \frac{t-t_0}{T} \right|^{p_{\beta}} +1 
\right)^2  + \hat{\rho}^2_2 \left(\left| \frac{t-t_0}{T} \right|^{p_{\beta}} + \hat{\rho}_3 
\right)^2 \right)^{\frac{1}{2}} }{ \left| \frac{t-t_0}{T} \right|^{p_{\beta}} + 1}  \\
a_4 &=& \sqrt{2} sign(\Omega) \arctan{\left( 
\frac{\left| \frac{t-t_0}{T} \right|^{p_{\beta}} + \Psi}{\left| \Omega 
\right|}  \right)} + a^0_4
\end{eqnarray}
Here we have defined the following constants.
\begin{eqnarray}
\Psi &=& 
\frac{\hat{\rho}^2_1 + \hat{\rho}^2_2 \hat{\rho}^2_3 }{\hat{\rho}^2_1 + \hat{\rho}^2_2} \\
\Omega &=& \frac{\hat{\rho}_1 \hat{\rho}_2 \left(\hat{\rho}_3 -1 \right)}
{\hat{\rho}^2_1 + \hat{\rho}^2_2} \\
\beta_0 &=& \log{\left( 2 q_{G5} \hat{\rho}^2_2 \left(\hat{\rho}_3 -1 \right)^2  \right)} 
\end{eqnarray}
The `expansion powers' (the $p$'s) are subject to the following constraint. 
\begin{eqnarray}
p_{\beta}^2 + \frac{1}{2} \left(p_{\varphi}^2 + p_{\beta_3}^2 \right) = \frac{2}{3}
\end{eqnarray}
We see that we end up with 11 independent integration constants, 
$p_{\varphi}$, $p_{\beta_3}$, $\hat{\rho_1}$,
 $\hat{\rho_2}$, $\hat{\rho_3}$, $T$, $t_0$, $\beta_3^0$, $\phi_0$, $a_4^0$ and $\alpha_0$. This is as 
it should be as we have solved 6 second order equations subject to one constraint.

\vspace{0.5cm}

\begin{figure}[t]\centering
\includegraphics[height=7.5cm,width=10cm]{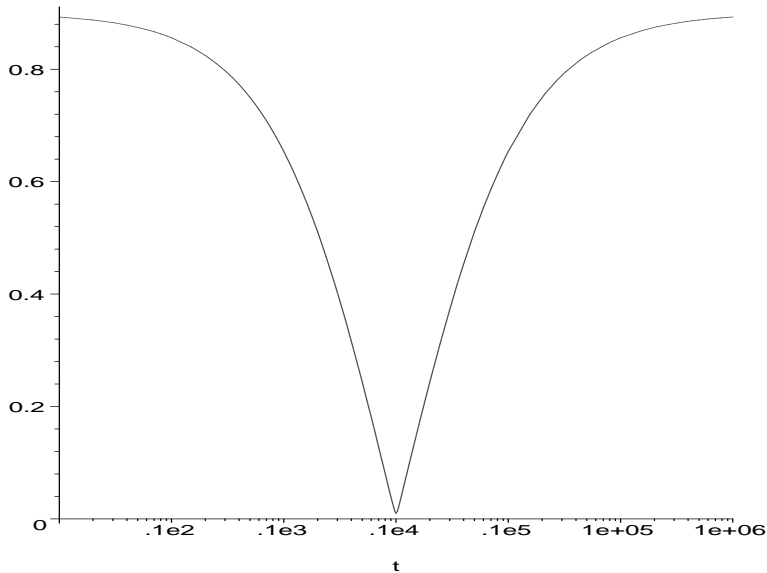} \vspace{0.25cm}
        \caption{\emph{The evolution of the size of the gauge field configuration relative to the 
transverse space, $\hat{\rho}$, is plotted against time. The effect of the `potential' generated for 
$\hat{\rho}$ by the 
motion of the embedding modulus, which is of the form $\frac{1}{\hat{\rho}^2}$, can clearly be 
seen. It prevents the instanton from reaching zero size. The asymptotic behaviour, where the 
modulus approaches a constant is also clearly visible. This plot corresponds to the following 
choices of integration constants: $T=10^3$, $p_{\b}=0.8$, $\hat{\rho}_1= 0.01$, $\hat{\rho}_2=0.9$, 
$\hat{\rho}_3=-1$ and $t_0=0$. }}
\label{fig1}
\end{figure}

There is one important 
aspect of the dynamics these solutions describe that we would like to emphasise 
before going on to an analysis of the features of these expressions that are important from the 
point of view of describing our phase transition. The size of the gauge 
field configuration is positive definite, $\hat{\rho} \geq 0$. The physical reason for this is easy to see.
If we consider the action \eqref{s4dtrunc} it is obvious that we can immediately integrate out 
the field $a_4$ by integrating its field equation once and substituting the result
 back into the action. When we do this we see that the kinetic term for $a_4$ turns into a 
potential term for $\hat{\rho}$ which goes as $\frac{1}{\hat{\rho}^2}$. Thus the motion of the 
embedding modulus creates an `effective potential' for the size modulus which becomes infinite as 
the size approaches zero. In fact since all of the other moduli associated with our instanton have
 the same coupling to $\hat{\rho}$ it is reasonable to assume that similar behaviour will be 
obtained in cases where we keep more of the bundle moduli. These observations have important 
consequences for early universe cosmology in heterotic models. 
They imply that, for a generic set of initial conditions, we would not 
expect small instanton transitions, where a gauge field size modulus goes to zero and the 
solitonic object is emitted from the fixed plane as an M5 brane (in strongly coupled language), to
occur in the course of the systems evolution. Only for very special configurations where only the 
size of the gauge fields is allowed to evolve are such transitions possible. Such special 
configurations would only be obtained from a subset of measure zero of the total space of 
initial conditions. For the case at hand 
an example of $\hat{\rho}$ 
evolution is plotted in figure \ref{fig1}. We can clearly see the size modulus `bouncing off' the 
potential at small values. 

Conversely it 
seems reasonable to conjecture that, assuming a similar phenomena does not occur for M5 brane 
moduli, a gauge field configuration appearing dynamically as a result of a small instanton 
transition would have to appear with all of the bundle moduli, except $\hat{\rho}$, constant if the 
process is possible at all. It should
 be noted that the behaviour we have discussed in the last few paragraphs 
is due to a property of the instanton moduli
 space rather than of the 
particular M-theoretic setting in which it is embedded. As such we would expect these observations
to hold for similar cases such as D3 branes dissolved on stacks of 
D7 branes and their 
associated small instanton transitions \cite{Johnson:gi}.

In short we conclude that it is extremely unlikely that M5 branes would have been 
created via small instanton
transitions in the early universe if it were described by the kind of compactification under 
consideration in this paper.

\vspace{0.5cm}

Let us now proceed to examine other aspects of 
our cosmological solutions is some more detail. As is usual for 
such systems, in order for the logarithms in our solutions to be well defined, the range of $t$ 
is restricted to one of two possibilities.
\be
 t\in\left\{\ba{clll}
       \left[ -\infty ,t_0\right]\;
,&(-)\;{\rm 
branch} \\
       \left[ t_0,+\infty\right]\; 
,&(+)\;{\rm branch}
       \ea\right.\; .
\ee
Our solutions therefore come in two types, positive and negative time branches. 
One can easily verify that the former starts
out in a past curvature singularity while the latter ends in a future
curvature singularity.

In either type of situation the solutions reduce asymptotically to rolling radii solutions 
\cite{Mueller:1989in,Brandle:2000qp}. For 
example in the positive time branch case as $t \rightarrow t_0$ the bundle moduli approach 
constant values and the exponentials of the rest of our fields exhibit power law behaviour, as 
they would if the bundle moduli were absent. The same happens as $t \rightarrow \infty$. The 
final expansion powers are the same as the initial ones with the exception of $p_{\beta}$ which 
is swapped in sign.

In between these two extremes something interesting happens. At some time, determined by the 
integration constant $T$, the initially constant $a_4$ moves swiftly and monotonically 
from its initial to its final value. The size modulus $\hat{\rho}$, at the same time $T$, 
also undergoes some non-trivial evolution to get between its two asymptotic values. It is the 
non-trivial motion of these two fields which causes the expansion power of the $\beta$ modulus 
to reverse and so this reversal naturally happens at the same time.

\begin{figure}[ht]\centering
\includegraphics[height=6.5cm,width=10cm]{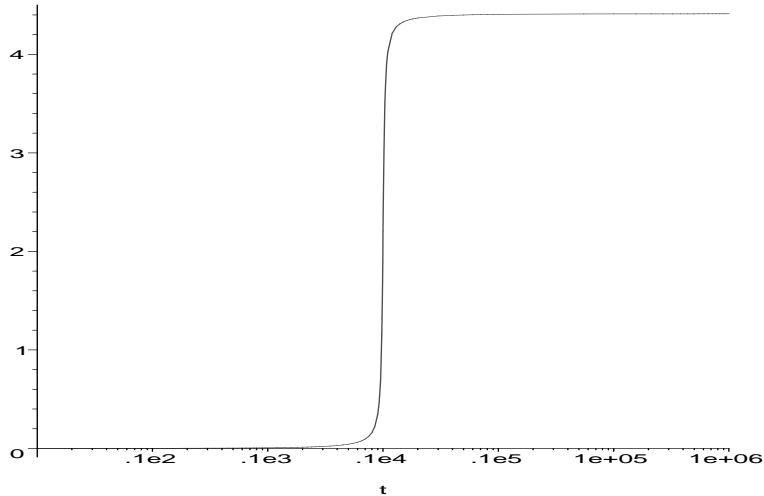} \vspace{0.25cm}
        \caption{\emph{The evolution of the embedding modulus, $a_4$, 
of the gauge field lump is plotted against time. The features 
described in the text - i.e. a constant embedding of the instanton within the gauge group 
at early and late times with a rapid monotonic change between the two embeddings at some 
intermediate time - are clearly visible. This plot corresponds to the same choice 
of integration constants as was made in figure \ref{fig1}. }}
\label{fig2}
\end{figure}

Concentrating on the embedding modulus we are left with the following picture. We start with 
a constant $a_4$ denoting a lump of $SU(2)$ valued gauge field which is sitting in some part 
of the larger gauge 
group and breaking a corresponding part of the symmetry. Then, at some time $T$, $a_4$
suddenly changes and the instanton's embedding within the gauge group rapidly changes 
to reach a final 
embedding where it again `comes to rest'. Thereafter the gauge field configuration 
is again sitting in some $SU(2)$ 
subgroup of the larger gauge group but this subgroup is embedded differently to the original one. 
Correspondingly the $SU(2)$ valued expectation value of the gauge field is now breaking a 
different part of the gauge symmetry. In other words at time $T$ there is a rapid change 
in the embedding modulus which results in a change in the unbroken gauge symmetry seen in four 
dimensions. An example 
of the evolution of $a_4$, for the same choice of integration constants as was used in figure 
\ref{fig1} is provided in figure \ref{fig2}. 

\vspace{-0.2cm}

In the next 
section we will explain how the dynamics we have just described 
can be incorporated into a phase transition where the 
rank of the unbroken gauge symmetry group is lowered during the transition (the reverse is 
also possible).

\begin{figure}[ht]\centering
\includegraphics[height=6cm,width=10cm]{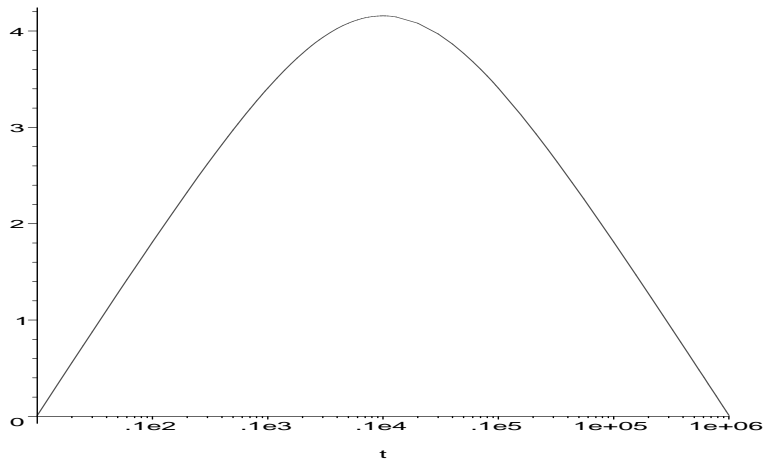} \vspace{0.25cm}
        \caption{ \emph{ The evolution of the size modulus associated with the space transverse to 
the gauge field configuration, $\beta$, is plotted against time. It is 
clear that the space collapses in size asymptotically as described in the text. This plot 
corresponds to the same choice of integration constants as was made in figure \ref{fig1}. }}
\label{fig3}
\end{figure}

Another point which should be mentioned is that, in the case of a positive time branch solution 
for 
example, the initial expansion power associated with $\beta$ is positive and the final one is 
correspondingly negative. This means that in these solutions the portion of the hidden space 
transverse to the instanton always collapses asymptotically (see figure \ref{fig3}). This should 
not worry the reader too much as there are many effects which will come in at late time to counter this 
problem. For example the kinetic energy of the fields redshifts with time due to the expansion of the 
universe. As such at some stage in the future evolution the potential terms which we have ignored will no 
longer be negligible when compared to the kinetic energy and the simple solutions presented here 
will no longer be valid.

\subsection{The phase transition.}

We are now ready to describe our explicit modulus driven, gauge symmetry breaking,
 phase transition.
The idea is to use the dynamics described in the previous subsection to rotate an instanton in 
one of these compactifications from an already broken part of the gauge group to an unbroken 
part. In other words we choose Wilson lines and the rest of the bundle such that the instanton
initially breaks part of the gauge group which is also 
broken by this other gauge field structure. We do this in such a manner as to allow the solutions 
presented in the previous subsection still to be valid - as we will describe in a moment. Then,
as the embedding modulus $a_4$ undergoes a change of value at time $T$, the instanton's 
embedding 
within the gauge group changes so that it breaks some previously unbroken part of the symmetry. 
Meanwhile all of the parts of the gauge group which were initially broken 
remain broken due to the Wilson lines and the rest of the bundle.

Such situations could be regarded as reasonably generic within these kind of compactifications. 
If we have an instanton which is initially breaking an already broken part of the gauge symmetry 
 then the only 
other possible dynamics are either that it is exactly stationary
or that it moves in just 
such a manner as to stay breaking a part of the gauge group that was already broken. In short we would 
expect the bundle moduli to drift around in value in the early universe (before moduli stabilisation) 
and they will generically change the four dimensional gauge group as they do so.

\vspace{0.5cm}

As an explicit example of the above, and one that is useful within the context of 
baryogenesis, we shall consider the following set up.

We split the gauge bundle into two pieces, a piece whose moduli we will consistently truncate
and a piece which will act as the instanton of our previous analysis. We shall choose the 
static part of the bundle to be $SU(4)$ 
valued and not overlapping in the internal 
space with  the instanton. Such a configuration can be created by combining several instantons identical 
to the one we have been considering rotated by elements of an appropriate $SU(4)$. We will 
fix the size of these instantons relative to the transverse space at 
sufficiently small values that it is possible for these objects to be localised away from our 
evolving gauge field configuration.
This part of the gauge bundle 
breaks the gauge symmetry seen in the four dimensional effective theory to $SO(10)$ in
the visible sector in the case of an $E8 \times E8$ ten dimensional gauge group.

We then add a Wilson line to the compactification which breaks the visible $SO(10)$ down to 
$G_{\textnormal{sm}} \times U(1)_{B-L}$ where $G_{\textnormal{sm}} = SU(3) \times SU(2) \times U(1)$ is 
the gauge group of the standard model.
We have denoted the last abelian subgroup $U(1)_{B-L}$ in a manner which is suggestive for the 
application of this work to baryogenesis. 
It is clearly possible to add such a Wilson line without altering the fact that our 
$SU(4)$ bundle is a supersymmetric solution to the equations of motion. It is also clear that the 
Wilson line does not alter the analysis of the dynamics of the moduli of the instanton which we 
are now going to add as it can be gauged  away on the finite volume where the instanton has 
appreciable support.

Finally we add a single $SU(2)$ valued instanton of the type we have been considering throughout 
this paper. We add this in such a way that it is initially embedded so as to break 
part of the gauge symmetry
which is already broken by the rest of the gauge bundle. 
We shall choose our integration constant $a_4^0$ of the previous section 
such that the initial value of the embedding modulus for which this is achieved is $a_4=0$. 
We also require that $a_4$, should it 
change, will rotate the gauge field configuration out of the $SU(4)$ subgroup and into a larger 
subgroup in such a manner 
as to break $U(1)_{B-L}$.

We are then, by using the dynamics described in the previous section, left with the following 
description of our phase transition. We start at some initial time, in a positive time 
branch solution, with the various metric moduli evolving in some rolling radius solution, with 
constant bundle moduli and with an unbroken visible gauge group of 
$G_{\textnormal{sm}} \times U(1)_{B-L}$. Then at 
some time, which is determined by the choice of integration constant $T$, the bundle moduli 
swiftly change. The embedding modulus $a_4$ in particular monotonically changes from its initial 
value to a final one. 
During this process the visible gauge group is broken to $G_{\textnormal{sm}}$. Finally the 
system reverts to a rolling radius type of evolution with its new gauge group. This then is our 
explicit description of a moduli driven phase transition.

In distinction to the other moduli driven phase transitions mentioned in the introduction 
we have an explicit dynamical description of this process. This means that we can write down, 
for example, how the mass of the 
gauge boson associated with the $U(1)_{B-L}$ factor changes with time during the phase transition. 

\vspace{1cm}

We shall now examine how the mass of the $U(1)_{B-L}$ gauge boson changes during the phase transition. 
The relevant mass term is generated solely by the terms proportional to 
 $\textnormal{tr} F_{A \; \mu} F^{A \; \mu}$ in the higher dimensional Lagrangian 
if we are working to first order in $\a'$.
When we substitute our vacuum solution and 
four dimensional fields into these terms we get contributions proportional to 
$\textnormal{tr} \left\{[A_{\mu}^{U(1)},A_B][A^{\mu \;U(1)},A^B] \right\}$, where $A_{\mu}^{U(1)}$
is the four dimensional gauge boson associated with the $U(1)_{B-L}$ factor of interest and $A_B$ is 
the background gauge field configuration which is our instanton. This will clearly give 
a mass term for the gauge boson in the four dimensional theory upon dimensional reduction. 
We find,
\begin{eqnarray}
S_{\textnormal{mass term}} = \frac{1}{2 \k_{4D}^2} \int d^4 x \; 18 \; q_{G5} e^{-\beta} \hat{\rho}^2 
\left(2 - \cos^4(a_4) - \cos^2(a_4) \right) A_{\mu}^8 A^{8 \; \mu}  \;\;\; .
\end{eqnarray}
Here we recall that $q_{G5} = \frac{\a' (2 \pi)^2}{V_{\textnormal{trans}}}$. 
We shall make a number of observations about this 
expression before proceeding. Firstly we note that the size of the mass 
is determined, in part, by a factor $\hat{\rho}^2 
\left(2 -\cos^4(a_4) - \cos^2(a_4) \right)$. The $\hat{\rho}^2$ part of this factor 
tells us that the mass of the $U(1)_{B-L}$ gauge boson is determined by the size of the gauge 
field configuration 
which breaks the relevant symmetry as compared to the size of the space transverse to that 
instanton. This is a sensible result from the point of view of considering what would happen in a 
small instanton transition. In such a transition the instanton would shrink down to zero size and 
then disappear from the vacuum to be replaced by an M5 brane (in the strongly coupled language). 
The M5 brane does not break the gauge symmetry in the same way that the instanton does and
so after our gauge field configuration is involved in such a transition we would expect the $U(1)_{B-L}$ 
gauge field to be massless. We see that the above result tells us this happens in a smooth and sensible 
manner.

The $\left(2 -\cos^4(a_4) -\cos^2(a_4) \right)$ factor in the mass term tells us that the mass 
of the gauge boson depends on the embedding of the instanton within the full higher dimensional 
gauge group. This is as it should be. For example if we take $a_4=0$ then the instanton is 
embedded in such a manner as to leave the $U(1)$ unbroken. As such the associated gauge boson 
should be massless and this factor ensures that this is indeed the case. In fact this factor can 
be thought of as a measure of how strongly the instanton configuration breaks the relevant $U(1)$.
As $a_4$ changes from zero the gauge group is broken more and more strongly 
and so the boson becomes more and more massive. The periodicity in $a_4$ merely reflects the 
periodicity of the $U(1)$ subgroup of $SU(3)$ with which this embedding modulus is associated 
when $a_1=a_2=a_3=0$.

Both of these above points raise a subtlety in deciding what is a light state in these kinds of 
compactifications. Following the normal approach in the literature the $U(1)$ gauge 
boson would be considered a heavy state and dropped from the four dimensional theory if the 
associated gauge symmetry was broken in the vacuum. We see from the above discussion that this is 
not necessarily a sensible thing to do. For example by making the transverse space to our 
instanton very large we can make $\hat{\rho}$ as small as we like while still keeping 
$\rho > \sqrt{\a'}$. As such we can make the mass of this supposedly heavy gauge boson as 
small as we desire. Similarly by choosing $a_4$ such that the relevant $U(1)$ is only very weakly 
broken by the instanton we can also make the associated gauge boson very light.

\vspace{0.2cm}

\begin{figure}[ht]\centering
\includegraphics[height=7cm,width=10cm]{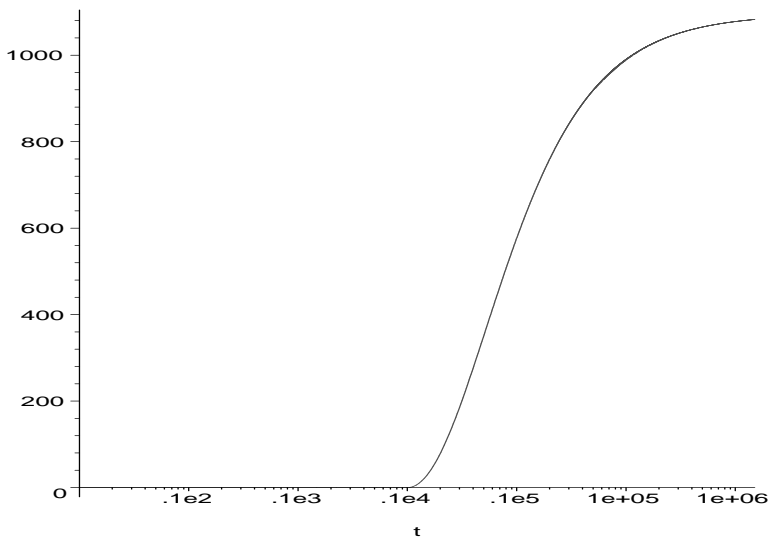}  \vspace{0.25cm}
        \caption{\emph{The evolution of the square of the mass of a gauge boson which is becoming heavy 
during the phase transition, relative to the square of the 
mass of a gauge boson which is merely a spectator 
to the process, has been plotted. The gauge field is massless before the phase transition when 
the relevant gauge symmetry is unbroken. Then during the phase transition the symmetry is broken 
and the gauge field obtains a mass. After the transition the gauge fields mass changes in time in 
the same way as all of the other massive bosons. As such the plot levels out to a constant value. 
This plot corresponds to the same choice of integration constants as was made in figure \ref{fig1}.
}}
\label{fig4}
\end{figure}

Let us now proceed to describe how the mass of this $U(1)$ gauge boson changes during our 
phase transition.
Due to the moduli describing the internal space changing with time even those gauge bosons which 
are massive at the start of the phase transition will have a mass which is evolving. 
In order to isolate the evolution of the mass of the $U(1)$ gauge boson which is due to the 
phase transition it is useful to remove the factors in the mass which are common to all of 
the massive gauge fields. This is done in the first of the plots below (\ref{fig4}). We can 
clearly see that the gauge boson is massless initially and remains so until the time of the 
phase transition (as determined by the integration constants of our solutions). 
Then it suddenly but smoothly increases 
to a non-zero value. At late time this plot levels off telling us 
that the phase transition has ended and that the newly massive gauge boson is now evolving in time
in the same way as all of the other massive gauge fields.

\vskip 0.4cm

\begin{figure}[ht]\centering
\includegraphics[height=7cm,width=10cm]{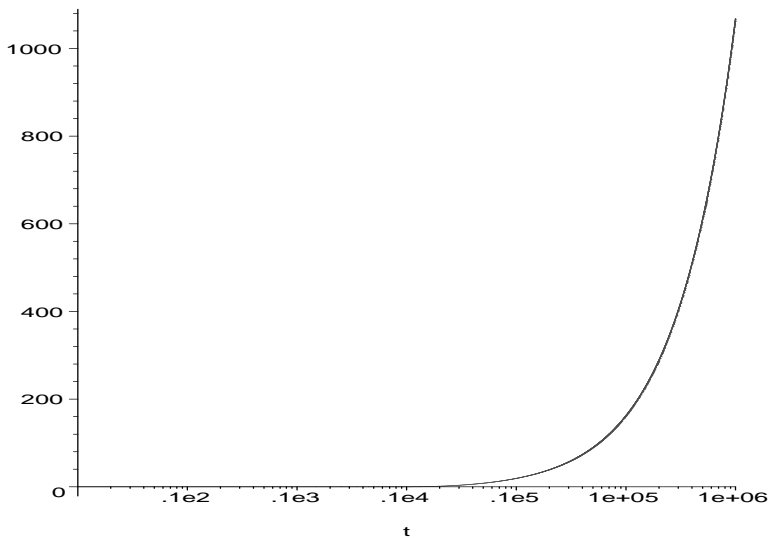}  \vspace{0.25cm}
        \caption{\emph{The evolution of the square of the absolute value of the mass of the $U(1)$ 
gauge boson involved in the 
phase transition with time. The mass diverges at late times as the transverse portion of the 
internal space collapses. This plot 
corresponds to the same choice of integration constants as was made in figure \ref{fig1}.}}
\label{fig5}
\end{figure}

If we put the full dependence on the internal space dynamics back in to our description we
find that we obtain the same sort of picture but with the mass increasing more rapidly after 
the collision and not leveling off at any stage (see figure \ref{fig5}). 
This is due to the fact, mentioned in the last 
section, that the size of the transverse part of the internal space is decreasing asymptotically.
\vskip 0.4cm

We could also examine how the mass of various matter fields, such as right handed neutrinos which 
would be of interest in applications to baryogenesis, change during the phase 
transition. The analysis would proceed in a very similar manner to that for the gauge boson 
present above. In particular the discussion presented in \cite{Lukas:1997fg} 
makes it very clear that the mass
term would come from a very similar source in the higher dimensional theory. That we 
can explicitly describe how these masses change with time is of course one 
advantage of being able to describe the phase transition from beginning to end in a well 
controlled way. This is in contrast to the use of a brane collision as a moduli driven 
phase transition.


\section{Conclusions.}
\label{conc}

In conclusion we have achieved the following in this paper. We derived the four dimensional 
effective theory for a class of heterotic models including various moduli which describe how the 
gauge field configuration in the vacuum changes with space and time. In particular we added to 
previous work in that we have included moduli which describe the embedding of an $SU(2)$ valued 
instanton, which is part of this gauge bundle, within the gauge group. We have noted an 
interesting property of the resulting theory; the K\"ahler potential is proportional to the logarithm
of the volume of the compact space, even when the bundle moduli are present. We then consistently 
truncated the resulting theory to obtain a four dimensional action suitable for an analysis of 
cosmological evolution.

We solved this four dimensional theory exactly to find the complete class of solutions based upon the 
FRW ansatz. In doing this, and by taking into account a few ancillary considerations, we were able to 
come up with the following explicit description of a moduli driven phase transition.
The phase transition begins with the system evolving in a rolling radius solution with the bundle 
moduli being constant. At some time, 
determined by one of the integration constants, the bundle moduli evolve from their initial to some 
final constant values. In particular an embedding modulus evolves in this manner indicating that the 
$SU(2)$ valued instantons embedding within the overall gauge group has changed. We chose this to occur 
in such a manner that the object 
is finally embedded so that it breaks 
a previously unbroken part of the visible gauge symmetry. Thus the system ends up in a final rolling radius 
solution with a new visible gauge group. We then emphasised the explicitly describable nature of this phase 
transition by  plotting the mass of the gauge boson 
associated with the symmetry broken in the transition against time. The mass was seen to smoothly change
from zero before the transition to a non-zero value after it. It should be emphasised that this phase 
transition is moduli driven and not thermal. There is no correlation between the mass of the particles 
which become heavy during the transition and the background temperature of the universe at the time 
at which the transition occurs.

Our cosmological solutions also showed that, for all but a set of measure zero of the initial 
conditions, small instanton transitions are classically 
forbidden during the course of the cosmological evolution of this system.

\section{Acknowledgments}

The author would like to thank Tim Hollowood for an extremely useful email, Andr\'e Lukas and 
Richard Ward for useful discussions and Douglas Smith for reading through the manuscript prior to its release. 
JG is supported by PPARC.
\section*{Appendix A. Index conventions.}

We use $\mu,\nu$, $A,B$, $M,N$ and $a,b$ as spacetime indices. The indices associated with the four 
non-compact dimensions are denoted $\mu,\nu = 0..3$. The indices $M,N=4,5$ are associated with the 
coordinates of the hidden space which our gauge field configurations are independent of. The 
remaining spacetime dimensions are denoted $A,B=6..9$. The indices $a,b = 4..9$ run over the entire 
compact space.

We use $u,v$, $\a,\b$ and $\dot{\a},\dot{\b}$ as ADHM indices 
(this is the same as the notation in \cite{Dorey:2002ik}).
 The indices associated with the fundamental of $U(3)$ are denoted $u,v=1..3$. The indices 
$\a,\b,\dot{\a},\dot{\b}=1,2$ are the usual two dimensional indices associated with the quaternionic 
notation used in, for example, \cite{Dorey:2002ik}. These indices are raised and lowered by the 
anti-symmetric tensor $\e$ where we take $\e_{21}=\e^{12}=1$. These two types of indices are often combined 
into a composite index, $(u+\a)$, which clearly then takes 5 possible values.

An index associated with a general modulus is written as $m$. We also use $m$ to denote a general modulus 
field itself.

A bar above an object denotes conjugation and in particular we use the conventions of \cite{Dorey:2002ik} 
if the object carries ADHM indices. In particular,

\begin{eqnarray}
\bar{\Delta}^{\dot{\a} (u+ \a)} = (\Delta_{(u + \a) \dot{\a}})^{*}
\end{eqnarray}

For the bosonic part of a chiral superfield the conjugation simply refers to complex conjugation in the 
usual manner.

\end{document}